# Ultrafast Internal Conversion in Ethylene. II. Mechanisms and Pathways for Quenching and Hydrogen Elimination

T. K. Allison,[1,2] H. Tao,[3,4] W. J. Glover,[3,4] T. W. Wright,[2,5] A. M. Stooke,[1] C. Khurmi,[2] J. van Tilborg,[2] Y. Liu,[1,2] R. W. Falcone,[1,2] T.J. Martínez,[3,4] and A. Belkacem[2]

[1]University of California at Berkeley, Berkeley, CA 94720
[2]Ultrafast X-ray Science Laboratory, Lawrence Berkeley National Laboratory, Berkeley, CA, 94720
[3]Dept. of Chemistry and the PULSE Institute, Stanford University, Stanford, CA 94035
[4]SLAC National Accelerator Laboratory, Menlo Park, CA
[5]Dept. of Chemistry, University of California at Davis, Davis CA, 95616

**Abstract:** Through a combined experimental and theoretical approach, we study the nonadiabatic dynamics of the prototypical ethylene $C_2H_4$ molecule upon $\pi \rightarrow \pi^*$ excitation with 161 nm light. Using a novel experimental apparatus, we combine femtosecond pulses of vacuum ultraviolet (VUV) and extreme ultraviolet (XUV) radiation with variable delay to perform time resolved photo-ion fragment spectroscopy. In this second part of a two part series, the XUV (17 eV < hν < 23 eV) probe pulses are sufficiently energetic to break the C-C bond in photoionization, or photoionize the dissociation products of the vibrationally hot ground state. The experimental data is directly compared to excited state *ab initio* molecular dynamics simulations explicitly accounting for the probe step. Enhancements of the $CH_2^+$ and $CH_3^+$ photo-ion fragment yields, corresponding to molecules photoionized in ethylene ($CH_2CH_2$) and ethylidene ($CH_3CH$) like geometries are observed within 100 fs after $\pi \rightarrow \pi^*$ excitation. Quantitative agreement between theory and experiment on the relative $CH_2^+$ and $CH_3^+$ yields provides experimental confirmation of the theoretical prediction of two distinct conical intersections and their branching ratio (Tao, et al. J. Phys. Chem. A. **113**, 13656 (2009)). Fast, non-statistical, elimination of $H_2$ molecules and H atoms is observed in the time resolved $H_2^+$ and $H^+$ signals.

# I. Introduction

Over the last decade, a picture of photochemistry has emerged in which conical intersections (CIs) between different electronic states provide an efficient "funnel" for non-radiative decay.[1-3] These CIs are not isolated points in configuration space but instead high-dimensional ($N$-2 dimensional, where $N$ is the number of internal degrees of freedom) seams. Thus, decay may occur at more than one location along the seam of CIs, so that one might best consider a set of mechanistic paths involving different classes of CIs. Few experiments have probed CIs directly, in part because the lifetime of a molecule in the region of a CI is generally extremely short – as little as a few femtoseconds.[4] Perhaps the most nearly direct measurement to date of dynamics around a CI is a recent experiment by Cerullo and coworkers.[5]

Before a photoexcited system can undergo internal conversion through a CI, the molecule must first rearrange to find it. One of the most ubiquitous mechanistic paths for non-radiative decay is through photoisomerization about carbon-carbon double bonds. For example, in the Xanthopsin and Rhodopsin protein families, the excitation of an electron in a conjugated $\pi$ system to a $\pi^*$ orbital "unlocks" the double bond, leading to torsion about the bond and allowing the chromophore to rapidly find the non-radiative decay path proceeding through conical intersections.[6,7] The simplest molecule with a carbon double bond is ethylene ($C_2H_4$), yet even this simple case exhibits rich internal conversion dynamics. As a prototypical system, the VUV excitation and subsequent photodissociation of ethylene have been the subject of extensive studies. Mulliken established[8] the labels $N$ for the ground state in which the $\pi$ orbital is filled ($\pi^2$) and $V$ for the valence excited state ($\pi\pi^*$). Also energetically in the vicinity of the $V$ state is a doubly excited state with zwitterionic character labeled $Z$ ($\pi^{*2}$). Ethylene's first absorption band, peaking at around 165 nm, is dominated by the $\pi \rightarrow \pi^*$ transition from the ground electronic $N$ state to the valence excited $V$ state.[9,10] In the ground state, the molecule is kept planar and rigid by the filled $\pi$ orbital, whereas in the excited state, it has a much lower potential energy in a twisted configuration. Because of ethylene's small size, the dynamics following $\pi \rightarrow \pi^*$ excitation have served as an important benchmark case for *ab initio* calculations using various techniques[11-15] that seek to predict the non-adiabatic dynamics of larger systems. The calculations predict that the molecule relaxes to the ground state $N$ through two general classes of conical intersections, one occurring near twisted and pyramidalized structures and the other near an ethylidene configuration ($CH_3CH$), where one of the hydrogen atoms has migrated across the double bond. The energy of the VUV photon is thus converted into vibrational energy on the ground state potential energy surface (PES).

Experimentally, the asymptotic outcomes of the dynamics have been studied in detail. The molecule is observed to eventually dissociate via two main channels: (1) eliminating a hydrogen molecule ($H_2$) or (2) eliminating two hydrogen atoms (2H). Lee and coworkers have conducted several studies of the kinetic energy distributions of the photodissociation products from ethylene and its di-deuterated forms (e.g. $CH_2CD_2$) after 157 nm excitation.[16-19] The studies of deuterated isotopomers illuminate pathways for hydrogen molecule elimination and provide indirect evidence for a fraction of the dissociating population proceeding through the theoretically predicted ethylidene configuration. For example, $H_2$ and $D_2$ molecules are observed after photoexcitation of *trans*-HDC=CHD. However, the differences observed between HD elimination from *cis* and *trans* forms of HDC=CHD show that the hydrogen atoms are not completely randomized.

The conical intersections and transition states between photo-excitation and dissociation can be observed with time resolved studies, providing important constraints for theory. However, ethylene's small size puts its excitation and photoionization energies in the VUV, beyond the spectral range of conventional femtosecond pump/probe techniques. As a result, it has been the subject of relatively little time resolved work. Previous work[20-23] has utilized multiphoton ionization (MPI) to investigate the lifetime of the excited *V* state, resulting in an apparent discrepancy for the excited state lifetime between theory and experiment. In part I of this series,[24] we reported the results of time resolved 7.7 eV pump/7.7 eV probe measurements, and showed that theory and experiment are in agreement when the decay of the photoionization amplitude is properly accounted for. In the present paper, we report the results of time resolved 7.7 eV pump/XUV (17-23 eV) probe experiments. The XUV photons are sufficiently energetic to break the C-C bond, allowing the arrangement of the hydrogens to be deduced from the ratio of symmetric ($CH_2^+ + CH_2$) to asymmetric ($CH^+ + CH_3$ or $CH + CH_3^+$) breakup of the ion, as illustrated in Figure 1. Furthermore, the XUV probe photon energy is also sufficient to photoionize $H_2$ molecules and H atoms, time-resolving the dissociation that occurs after internal conversion. As in part I, we simulate the experiment using the *ab initio* multiple spawning method (AIMS). In addition to calculating the time-dependent photoionization amplitude of the excited neutral molecule, fragmentation of the ion is also simulated by performing dynamics simulations on the cation surfaces after photoionization, allowing a direct comparison with experiment.

**II. Methods**

The apparatus is depicted in Figure 2 and described in greater detail in previous reports.[25,26] High order harmonics of 807 nm are generated with a repetition rate of 10 Hz by loosely focusing (f = 6 m) 30 mJ, 50 fs laser pulses into a 5 cm gas cell with laser drilled pinholes. The cell is filled with 2.4 Torr of Xe and scanned through the focus to optimize the harmonic yield. Greater than $10^{10}$ photons/harmonic/shot emerge from the gas cell in orders 11-15 (hv = 17-23 eV) along with more than 200 nJ in the fifth harmonic (hv = 7.7 eV).[26] The harmonic and fundamental beams are allowed to diverge for three meters where they are incident on a super-polished silicon mirror set at the 800 nm Brewster angle (75°). The silicon mirror removes the fundamental and reflects the harmonics.[27]

Pump/probe delay is achieved with a split mirror interferometer (SMI). The harmonics are focused into a pulsed molecular beam of neat ethylene by two "D-shaped" spherical concave mirrors (r = 20 cm) at normal incidence. Photo-ions from the focal region are selected with a 500 µm aperture and measured with a time of flight ion mass spectrometer (TOF). We checked that the results reported here are not influenced by dimers in the molecular beam by varying the backing pressure of the pulsed valve.

The probe arm mirror is mounted on a piezoelectric translation stage to produce a time delay. Wavelength selection in each arm of the SMI is achieved through a combination of transmission filters and coatings on the two D-shaped mirrors. In the probe arm, we inserted a thin tin (Sn) foil to select harmonics 11-15. The probe arm mirror consisted of a magnetron sputtered 30 nm $B_4C$ coating (R ≈ 30% for hν = 17-25 eV) on a fused silica substrate. The pump arm mirror was a multilayer dielectric stack (Layertech Gmbh) designed to reflect and focus the fifth harmonic (R>90%) and transmit the 3rd harmonic and fundamental. We also inserted a 230 µm thick $CaF_2$ window in the pump arm which temporally separates the 5th harmonic from residual 3rd harmonic and fundamental light.[28] Thus, residual 3rd harmonic and fundamental light pulses arrive at the focus more than 300 fs earlier than the 5th harmonic, and being non-resonant with the target $C_2H_4$ gas, pass through unabsorbed and do not corrupt the experiment. We confirmed this by scanning the pump/probe delay to the large negative delays where the 3rd harmonic and fundamental would be coincident with the probe. No effects on the photo-ion yields were observed. The mirrors and filters were moved perpendicular to the beam path to adjust the power in pump and probe arms. We varied the fraction of molecules excited in the focal volume (as measured by depletion of the $C_2H_4^+$ signal) between 3% and 9%. We also performed experiments where we replaced the Sn foil with an oxidized Al foil to shift the XUV probe photon spectrum to higher energies. The shape of the time resolved photo-ion signals reported here was found to be independent of the excitation fraction and the metallic foil used in the XUV probe arm. The time resolution of the apparatus was determined by measuring the ultrafast dissociation of water vapor as described previously.[25] The resulting finite instrument response function (FIR) is taken as a Gaussian of $44^{+15}_{-10}$ fs FWHM.

As in part I of this series, *ab initio* multiple spawning (AIMS) molecular dynamics simulations were carried out to interpret the experimental results. Details of the AIMS method can be found in part I and previous publications.[11,29] Briefly, the wave function ansatz is written as superposition of both electronic and nuclear wave functions. The nuclear wave function is a linear combination of time-dependent trajectory basis functions (TBFs), which are sampled in accord with a v=0 Wigner distribution on the ground electronic state, in the harmonic approximation. These TBFs are propagated according to classical equations of motion.[30] New basis functions are added adaptively (spawned) when the total wave function moves into a region of large nonadiabatic coupling, e.g. near a conical intersection. The population transfer between TBFs is calculated according to the time-dependent nuclear Schrödinger equation in the time-evolving basis set of TBFs. The electronic structure problem is solved "on the fly" with the nuclear dynamics to determine the potential energy surfaces, interatomic forces, and nonadiabatic couplings as needed.

The primary observable in the present experiments is the distribution of measured fragment ions as a function of the pump-probe time delay. Thus, not only must the excited state (and, through nonadiabatic transitions, also the ground state) dynamics of the neutral molecule be simulated, but also the probability of ionization to each of the energetically accessible cation states must be determined and then the fragmentation outcome of molecules ionized to each of these cation states needs to be established. Due to the large internal energy of the system gained from the pump pulse, the simulated photoion fragment signals are not dominated by contributions from a single cation state. However, the distribution of cation surfaces populated is not uniform and does depend on the geometry and electronic wave function of the neutral molecule when it absorbs the probe pulse photon. A realistic projection of the photoionization amplitude onto each of the cation states is thus essential for initiating the cation dynamics. An experimental pump-probe study of cation dynamics[43] found the dynamics to be largely concluded within 200 fs. We thus conducted dynamics simulations on the cation surfaces for 200 fs after photoionization. Greater than 90% of the population on the cation surfaces has fragmented within this time window.

The dynamics on the neutral surface were calculated with AIMS using an SA-2-CAS(2/2)-MSPT2 wavefunction, as described previously.[24,29] Briefly, this nonmenclature means that two singlet states are included in a state-averaged complete active space self-consistent field wavefunction with an active space of two electrons and two orbitals. The wavefunction is further corrected using the multi-state variant of second-order perturbation theory (MSPT2) in order to include dynamic electron correlation effects. All calculations use the 6-31G* basis set, unless otherwise specified. The photoionization probability was calculated using SA-5-CAS(6/5) and SA-5-CAS(5/5) for the neutral and cation electronic wavefunctions, respectively. Dynamics of the cation after photoionization was computed using the SA-5-CAS(7/6) electronic wavefunction and includes nonadiabatic transitions among the cation states (as does the neutral dynamics prior to photoionization). The increased active space for the calculation of photoionization probabilities and cation dynamics allows the modeling of the lowest five cation states and the description of fragmentation processes. The photoionization probability is calculated using a first-order Born approximation, as discussed in detail in part I of this series.[24] The photon energy used for the calculation of photoionization probabilities is 20eV (midway in the range of XUV harmonics comprising the probe pulse[31]). We do not attempt to calculate absolute photoion yields (which would require detailed information about the pump and probe intensities). Thus, the theoretical results are scaled by a common scale factor for comparison with experiment. The photoionization

Allison, et al. – VUV/XUV Spectroscopy of Ethylene – Page 7

probabilities are averaged over all possible orientations of the molecule with respect to the laser polarization.[32]

The calculation of the photo-ion fragment yields proceeds as follows. First, the calculation of the dynamics for the excited neutral molecule is carried out as discussed above and in part I. Then, the total signal at a given pump-probe time delay is written as a sum of contributions from each of the cation states:

$$S_{frag}(t) = \sum_{\alpha}^{M_{cation}} S_{frag}^{\alpha}(t) \tag{1}$$

where $M_{cation}=5$ (cation states labeled X through D in previous work[33,34]) is the number of cation states that are considered for photoionization, $\alpha$ is the adiabatic index of the cation state, $t$ is the pump-probe delay, and *frag* denotes one of the monitored fragment ions ($CH_2^+$, $CH_3^+$, $H^+$, and $H_2^+$). The contribution for each cation state is calculated as:

$$S_{frag}^{\alpha}(t) = \frac{1}{N_{initial}^{neutral}} \sum_{i}^{N_{neutral}(t)} n_i(t)\sigma_{i\alpha}(t) \sum_{j}^{N_{cation}(t'=200\,fs)} n_j^{cation-\alpha}(t'=200\,fs)\delta_{frag(j\alpha),frag} \tag{2}$$

where $N_{neutral}(t)$ and $N_{cation}(t'=200fs)$ are the number of TBFs representing the neutral molecule at the given pump-probe time delay and the number of TBFs representing the XUV-ionized molecule at the end of the cation dynamics, respectively.[35] The population of the *i*th TBF at pump-probe time delay $t$ on a neutral state is denoted $n_i(t)$ and $\sigma_{i\alpha}(t)$ is the probability of ionization to cation state $\alpha$ from this TBF at this pump-probe time delay (this probability depends on the electronic state of the neutral TBF, but that dependence is suppressed in this notation for simplicity). For each cation state $\alpha$ with nonvanishing $\sigma_{i\alpha}(t)$, dynamics starting from the *i*th neutral TBF (beginning on cation state $\alpha$) is followed for 200fs (time in the post-probe dynamics is labeled as $t'$).[36] Then, each of the $N_{cation}$ TBFs ($N_{cation}$ may be more than one both because there are multiple cation states initially populated by the probe and also because of spawning during the post-probe cation dynamics) is analyzed to determine the identity of the resulting fragments, denoted as *frag(jα)* for the *j*th cation trajectory from photoionization to cation state α. We used a connectivity matrix method[37] in order to classify the fragments (see supporting information for details[38]). Two atoms are considered connected if the bond length between them is less than a constant multiple (1.5) of the sum of the covalent radii of the two atoms involved (0.700 Bohr for H atom and 1.285 Bohr for C atom). A matrix is formed with rows and columns denoting the atoms in the molecule and with bond order values for connected atoms and zero entries otherwise. This matrix is then related to standard connectivity matrices allowing for permutation of atoms with the same atomic number in order to determine the final fragments. We further assign the charge of each fragment by analysis of the electronic wavefunction. The probability for this fragment is then multiplied by the population of the *j*th cation trajectory from photoionization to cation state α at the end of the cation dynamics: $n_j^{cation-\alpha}(t'=200\,fs)$. Finally, the total signal is normalized by the number of initial conditions used to represent the neutral excited state dynamics, $N_{initial}^{neutral}$. The total number of TBFs that are propagated in this simulation, including dynamics on both neutral and cation states and TBFs created through spawning is in excess of 24000.

TBFs on the neutral state that have undergone nonadiabatic transitions to the ground electronic state often dissociate (independent of the XUV probe pulse). When this happens (or when highly distorted geometries are reached which can cause problems for the electronic structure in the limited active space we are using), we terminate the dynamics for this TBF, but account for its contribution to photofragments by assuming that the photoionization cross-section and fragment distribution ceases to evolve for times after dissociation.

Allison, et al. – VUV/XUV Spectroscopy of Ethylene – Page 9

**III. Results**

The experimentally measured differential photo-ion yields produced by the XUV probe pulse as a function of pump-probe delay are shown in Figure 3. Count rates of 0 to 3 ions per shot were recorded for $CH_2^+$, $CH_3^+$, $H_2^+$, and $H^+$ fragments using a threshold discriminator and a time to digital converter. The signal level and error bars were corrected for coincident counts assuming Poisson statistics. The $C_2H_4^+$ ion signal rate was roughly 80 ions/shot and was recorded by integrating the pulse area in the time of flight spectrum with a boxcar integrator. The fractional change in the $CH_2^+$ and $CH_3^+$ fragments is on the order of 10%, consistent with the observed depletion of the $C_2H_4^+$ signal.

The $C_2H_4^+$ fragment signal promptly decays as the molecule moves away from the Franck-Condon region and the overlap with bound states of the cation decreases. At short times after excitation, peaks arise in the $CH_2^+$ and $CH_3^+$ yields at ≈60 and ≈90 fs delay, respectively. The $H^+$ signal rises abruptly near time zero, while the $H_2^+$ signal monotonically rises on a much slower time scale.

The AIMS simulations of the excited state dynamics consist of 44 initial TBFs (randomly sampled from the Wigner distribution as discussed above) on the neutral excited *V* state. For each of these initial TBFs, the dynamics is followed for 600fs. This 600 fs simulation time is sufficient to ensure that all population has returned to the ground electronic state. The fragment yield due to XUV ionization is calculated as described above (for pump-probe time delays in intervals of 10 fs).

*IIIa. $CH_2^+$ and $CH_3^+$ Fragment Channels*

A ground-state bleach contribution is evident at long time delays in the fragment yield for $CH_2^+$ and $CH_3^+$ ions. This arises because not all the molecules in the focal volume of the SMI are excited by the pump pulse, and unlike most time resolved photoionization studies employing probe pulses with lower photon energies, the XUV probe pulse induces photoionization from *both* the molecules in the focal volume that absorb a photon from the pump pulse and those that do not. The AIMS simulations compute only the contribution from molecules excited by the pump pulse and therefore we must estimate and subtract the ground state bleach component to make a direct comparison. We do this for $CH_2^+$ and $CH_3^+$ in the following way. First, the magnitude of the bleach component at long time is taken from the average of the long delay data points of the signals. This magnitude multiplies a step function centered at zero time delay (zero for negative time and negative one for positive time), which is further convolved with a 44fs FWHM Gaussian representing the finite instrument response. The resulting bleach component is shown as a black dashed line in the $CH_2^+$ and $CH_3^+$ panels of Figure 3. The systematic uncertainty in the instrument response and statistical uncertainty in the fragment yield at long time delays are used to obtain a 1σ confidence interval of the bleach functions, indicated by the shaded gray regions in the $CH_2^+$ and $CH_3^+$ panels of Figure 3. For $H^+$ and $H_2^+$, the bleach contribution is negligible because the contribution to these signals from unexcited molecules in the focal volume is negligible.

The AIMS predictions (convolved with the 44 fs FIR) for the $CH_2^+$ and $CH_3^+$ fragment yields as a function of pump-probe time delay are shown in the upper two panels of Figure 4. The predicted ion yields are compared directly to the experimental data (after subtraction of the ground state bleach as discussed above) in Figure 5. Because only a finite number of initial TBFs are used in the AIMS simulations, there is an associated statistical uncertainty. This was estimated using bootstrap analysis[39] and the 1σ uncertainty is shown as shaded regions in Figure 5. The uncertainty in the bleach function (see Figure 3) has been propagated to calculate the experimental error bars in Figure 5, which become substantially larger than those of the raw data for points near time zero.

We emphasize that the theory curves in Figure 5 are not independently scaled to match the experimental data, but are both multiplied by the same scaling factor. In addition to agreement on the approximate shapes of the time dependent signals, theory and experiment then also agree on the relative magnitudes of the $CH_2^+$ and $CH_3^+$ signals. We note that the agreement on the relative peak heights is not critically dependent on the choice of the bleach functions, as can be seen from the y-axes of Figures 3 and 4. We also note that the measured ratio of $CH_2^+$ to $CH_3^+$ produced by the XUV probe at negative delays or with the pump pulse blocked agrees with that measured for the cold ground state by Ibuki et al.[33] for our probe photon energy band.

It is also of interest to determine how the ground and excited states of the neutral molecule contribute to these fragment yields. In Figure 4, we also show the fragment ion yields for $CH_2^+$ and $CH_3^+$ decomposed by the electronic state of the neutral molecule at the time of photoionization. The $CH_2^+$ fragment signal arises predominantly from photoionization of neutral molecules on $S_1$, especially at early times. For later pump-probe time delays, the observed $CH_2^+$ fragments arise from $S_0$. This is not surprising since the excited state lifetime of neutral ethylene is quite short (< 100fs). More interesting is the same analysis for $CH_3^+$ fragments, also shown in Figure 4. The initial fragments arise from $S_1$ at the earliest times, but then from $S_0$ and at even later times again from $S_1$. This suggests that the TBFs that give rise to $CH_3^+$ fragments are in the vicinity of a conical intersection – one might suspect this to be the ethylidene conical intersection seam given the fragmentation pattern. In order to investigate this further, we have taken all neutral TBFs which give rise to prompt (within the 200fs cation state simulation time) fragmentation after XUV photoionization and separated them according to the calculated fragment ($CH_2^+$ or $CH_3^+$). The geometries at the time of photoionization are aligned (by rotation, translation, and permutation of identical atoms as necessary) and the resulting density is plotted as a wireframe isosurface in Figure 6. For illustration, we show these densities superimposed on the twisted-pyramidalized and ethylidene-like MECIs of ethylene (represented as licorice structures). There is a clear correlation in both cases – neutral geometries which give rise to prompt production of $CH_2^+$ on XUV absorption are very close to the twisted-pyramidalized MECI and those which give rise to prompt production of $CH_3^+$ are very close to the ethylidene MECI. The fragment signals in the present VUV-XUV experiment are capturing the ethylene molecule while it is in the region of the conical intersection seams. The relative magnitudes of the $CH_2^+$ and $CH_3^+$ signals thus provide a measure of the relative importance of the two classes of conical intersections in ethylene.

*IIIb. $H_2^+$ and $H^+$ Fragment Channels*

The AIMS results for the cumulative H and $H_2$ elimination fractions convolved with the Gaussian FIR are shown in the lower two panels of Figure 4. The bootstrap-estimated 1σ statistical uncertainty is indicated by the shaded regions in the lower two panels for the H and $H_2$ elimination channels. These cumulative elimination fractions represent the percentage of the excited molecules that have dissociated to give H atom or $H_2$ molecule fragments before the arrival of the pump pulse. Thus, this is not directly comparable to the experimental signal, which includes the $H/H_2$ photoionization cross-section and possibly also $H^+/H_2^+$ from primary or secondary fragmentation after photoionization. We did not observe any $H^+/H_2^+$ production during the 200fs of propagation on the cation surfaces after photoionization, but we cannot preclude the formation of these ions on a longer time scale. Nevertheless, the qualitative features of the experimental $H^+$ and $H_2^+$ fragment yields shown in Figure 3 are well reproduced by the reported elimination fractions from the simulations.

As the asymptotic products of the photolysis of ethylene at 157 nm have been studied extensively,[16-19] we can make a quantitative comparison between the measured signals at our maximum delay of 600 fs and what would be expected if dissociation were complete. The dissociation channels and their branching ratios are:[18]

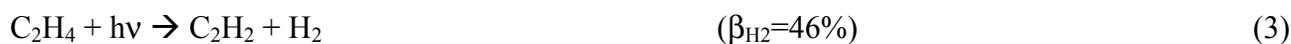

$C_2H_4 + h\nu \rightarrow C_2H_2 + H_2$    ($\beta_{H2}$=46%)    (3)

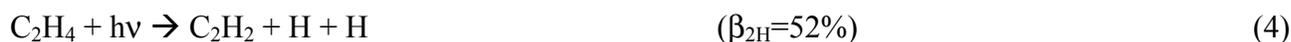

$C_2H_4 + h\nu \rightarrow C_2H_2 + H + H$    ($\beta_{2H}$=52%)    (4)

$$C_2H_4 + h\nu \rightarrow C_2H_3 + H \quad (\beta_H=2\%) \qquad (5)$$

The fraction of molecules in the focal volume excited is estimated from the depletion of the $C_2H_4^+$ signal to be 9%. We assume that $C_2H_4^+$ molecules are collected from the SMI focus with 100% efficiency through the 500 μm aperture at the entrance of the ion TOF. The emission of the photoproducts is taken to be isotropic, as the departure from isotropy is negligible for the present purposes.[19] The expected $H^+$ and $H_2^+$ signals are then given by:

$$\Delta S_{H_2^+} = -\Delta S_{C_2H_4^+} \frac{\beta_{H_2}\sigma_{H_2}}{\sigma_{C_2H_4^+}} CE[P_{H_2}(T)] \qquad (6)$$

$$\Delta S_{H^+} = -\Delta S_{C_2H_4^+} \frac{(2\beta_{2H}+\beta_H)\sigma_H}{\sigma_{C_2H_4^+}} CE[P_H(T)] \qquad (7)$$

where $\Delta S_{C_2H_4^+}$ is the change in the $C_2H_4^+$ rate in ions/shot, $\sigma_{H2}$ ($\sigma_H$) is the photoionization cross section of $H_2$ (H), and $CE[P(T)]$ is the collection efficiency for molecules with kinetic energy distribution $P(T)$. The branching ratios $\beta_i$ are given in equations 3-5. We estimate that the collection efficiencies for $H_2$ molecules and H atoms with the kinetic energy distributions $P(T)$ reported by Lin et al.[17] to be 59% and 98%, respectively. The drop in the $C_2H_4^+$ ion rate is estimated to be $\Delta S_{C_2H_4^+} \approx -0.09 \times 80 = -7.2$ ions/shot. Inserting these parameters into equations 6 and 7 gives the predicted values $\Delta S_{H_2^+} = 0.9$ ions/shot and $\Delta S_{H^+} = 1.0$ ions/shot. This estimate ignores the unknown contribution of $H^+$ and $H_2^+$ signals originating from dissociative photoionization of the $C_2H_2$ photoproduct. The $C_2H_2$ contribution is negligible if the cold ground state cross sections[40] are assumed, but may be significant considering the large internal energy. Systematic errors, such as misalignment of the ion TOF aperture with respect to the SMI focus would reduce these estimated yields. The values measured at long delay were $\Delta S_{H_2^+} = 0.34$ ions/shot and $\Delta S_{H^+} = 0.65$ ions/shot. This analysis of the data then leads to estimates of $H_2$ elimination being 0.34/0.9 = 38% complete and H elimination being 0.65/1.0 = 65% percent complete after 600 fs.

The extent to which dissociation is complete in the AIMS simulations can also be evaluated by looking at the fraction of the total population in dissociated trajectories. We find this to be 16% of the total population for $H_2$ elimination and 25% of the total population for H elimination. Comparing to the branching ratios of reactions 3-5, this corresponds to $H_2$ elimination being (0.16 ± 0.04)/0.46 = 35% ± 9% complete and H elimination being (0.25 ± 0.09)/0.54 = 46% ± 17% complete after 600 fs. The status of the second atomic H elimination event described in reaction (4) is unknown in the simulations because TBFs are terminated after the first dissociation (the active space used would need to be larger to provide a meaningful description of multiple dissociative processes).

**IV. Discussion**

The existence of a hydrogen migration channel ($H_3CCH$) channel in the photolysis of ethylene was originally proposed by Okabe and McNesby[41] in 1962 and has been supported by many molecular dynamics,[20-23] potential energy surface,[42] and statistical[43,44] calculations. Its direct experimental detection, however, has remained elusive. Lee et al. observed[18] the dissociation channel $CH_2CD_2$ → $C_2H_2$ + D + D, which should be energetically forbidden if both D atoms come off the same side of the molecule and the $C_2H_2$ molecule is left in the energetically higher vinylidene configuration. It can be concluded from this that a hydrogen migration must then occur before the second D atom is eliminated, but it is unclear whether it occurs before the first. In a time resolved multiphoton ionization study,[20] Kosma et al. observe long lived (> 1 ps) weak tails (< 1% of the peak height) on the $C_2H_3^+$ and $C_2H_2^+$ when the intensity of multiphoton ionization probe light is increased. They argue that the difference in these $C_2H_3^+$ and $C_2H_2^+$ signal tails necessitates the inclusion of (at least) two extra fit parameters in the (at least) 14 parameter fit used to describe the $C_2H_3^+$ signal, and that the inclusion of an additional time constant in the $C_2H_3^+$ fit implies the decay of two distinct configurations of the ground state. They assign these configurations to ethylene and ethylidene configurations, based on previous theoretical work. In the present work, the only assumption required for the assignment of the peak in the $CH_3^+$ signal to the ethylidene configuration is that the fragmentation along the C-C bond of the excited ion is prompt with respect to hydrogen migration. This is the case for photoionization of cold ground state ethylene. Although hydrogen migration can occur in the cation,[45] the $CH_2^+$ yield upon photoionization with the XUV probe light is roughly 10 times larger than that of $CH_3^+$, observed with the present apparatus and also previously.[33,34] Prompt C-C bond cleavage and the correlation of the $CH_2^+$ and $CH_3^+$ signals with ethylene and ethylidene geometries are also fully supported by our *ab initio* molecular dynamics simulations.

If the assignment of the $CH_3^+$ signal to the ethylidene configuration in the neutral is accepted, the observed time scale for hydrogen migration can be compared to previous theoretical results in addition to the ones presented here. Several molecular dynamics calculations have been published showing the time scale for hydrogen migration in ethylene. In a series of semi-empirical molecular dynamics simulations,[13-15] Barbatti et al. find significant population reaching hydrogen migrated geometries between 60 and 100 fs. In a previous AIMS study, Levine et al. show[46] the population on the *V*-state PES at twist angles around 90° and high pyramidalization angles greater than 120° peaking between 50 and 100 fs. A high pyramidalization angle indicates hydrogen migration across the double bond. These time scales are in good agreement then with the broad peak in the $CH_3^+$ signal occurring between 50 and 100 fs.

Dynamics simulations have also attempted to determine the percentage of population transfer from *N* to *V* occurring through twisted-pyramidalized geometries vs. geometries involving hydrogen migration. In earlier AIMS simulations using the CASSCF formalism for the electronic structure, Levine et al. found roughly equal fractions of the total populations decaying through the two CIs.[12,46] This was also observed in a semi-empirical molecular dynamics studies by Barbatti et al.[13] In the present AIMS simulations using MS-CASPT2 for the electronic structure, we find 72% of the non-radiative decay events occur through twisted-pyramidalized CIs, and 28% occur through ethylidene-like CIs. Given the agreement of the $CH_2^+/CH_3^+$ relative peak height and the higher accuracy of the MS-CASPT2 electronic structure methods used to determine the potential energy surfaces during the dynamics, we feel that the present prediction is more trustworthy than previous results.

The experimental and theoretical data support assigning the bulk of the $H_2^+$ signal to photoionization of hydrogen molecules during and following dissociation. Experimentally, four features of the $H_2^+$ signal support this assignment: (1) within the statistical error, it is monotonically increasing, (2) the monotonic rise coincides with the birth and decay of population near conical intersections, (3) the production of charged $H_2^+$ fragments from the photoionization of cold ground state ethylene is observed to be extremely small (<0.07 Mb) below the double ionization threshold,[33] and (4) the comparison of the observed $H_2^+$ increase with what would be expected from the asymptotes indicates a large partial photoionization cross section of the system for $H_2^+$ production, comparable to that of an $H_2$ molecule. For the $H^+$ data, observation (4) is fulfilled but observations (1-3) are not. Comparing the experimental data with the AIMS simulations, the $H^+$ signal rises ≈50 fs earlier than dissociation in the AIMS simulations. A large portion of $H^+$ signal at early times after excitation likely comes from a combination of photoionization of eliminated H atoms and dissociation in the cation after photoionization, whereas at later delays, the signal is likely dominated by photoionization of eliminated H atoms. At 600 fs pump-probe delay, the dissociation observed in the AIMS simulations and that estimated from the experimental data in section III are in reasonable agreement for both $H_2$ and H elimination.

The observed time scales for H and $H_2$ elimination observed in this work can be compared to Rice-Ramsperger-Kassel-Marcus (RRKM) calculations regarding the photolysis of $C_2H_4$,[43,44] which model dissociation of the hot ground state assuming a statistical redistribution of the photoexcitation energy. These calculations attain qualitative agreement with the experimentally measured[17] branching ratios, but significant discrepancies appear upon close comparison.[18] The calculated RRKM rate constants predict total time constants > 1 ps for $H_2$ and H eliminations, much slower than observed here. However, the calculated RRKM rate constant for $H_2$ elimination from ethylidene geometries is calculated to be much faster, in the range of $9.6 \times 10^{12}$ to $4.8 \times 10^{13}$ s$^{-1}$, with the rate-limiting step for this channel being the formation of the ethylidene transition state. The present work establishes the formation of ethylidene within 100 fs after photo-excitation. $H_2$ elimination can then proceed rapidly from the ethylidene geometry, as predicted by the RRKM studies and as also seen in the present AIMS simulations.[29]

**V. Conclusions**

We have employed a novel VUV/XUV pump-probe apparatus and the AIMS method to study the dynamics of the prototypical ethylene molecule upon $\pi \rightarrow \pi^*$ excitation. In the first paper of this series,[24] we resolved the apparent discrepancy between theoretical simulations and time resolved multiphoton ionization experiments for the excited state lifetime. In the present paper, the location of the hydrogen atoms is probed by breaking the C-C bond via XUV photoionization, providing direct evidence for the involvement of an ethylidene ($H_3CCH$) structure. The agreement between theory and experiment for the ratio of $CH_2^+$ to $CH_3^+$ peak heights and their relative positions in the pump-probe traces provides a strong confirmation of the theoretical predictions of two distinct non-radiative decay paths, one characterized by twist and pyramidalization and the other characterized by hydrogen migration, and their relative branching ratio. The time resolved $H^+$ and $H_2^+$ signals add new and important data to the already extensively studied, yet still unresolved problem of the photodissociation of $C_2H_4$. The experimental data and the AIMS simulations indicate the presence of fast, non-statistical, elimination channels for $H_2$ molecules and H atoms in the photolysis of $C_2H_4$.

**VI. Acknowledgements**

This work was supported by the US Dept. of Energy Office of Basic Energy Sciences, under contracts No. DE-AC02-05CH1123 and DE-FG-52-06NA26212. Theory work was supported under DOE Contract No. DE-AC02-7600515. A.M. Stooke gratefully acknowledges the full support of the Fannie and John Hertz Foundation. We thank A. Stolow, and C. Evenhuis for discussions and help with the code to calculate the connectivity matrix. We acknowledge C. Caleman, M. Bergh, H. Merdji, and M. P. Hertlein for help with the apparatus in its early stages.

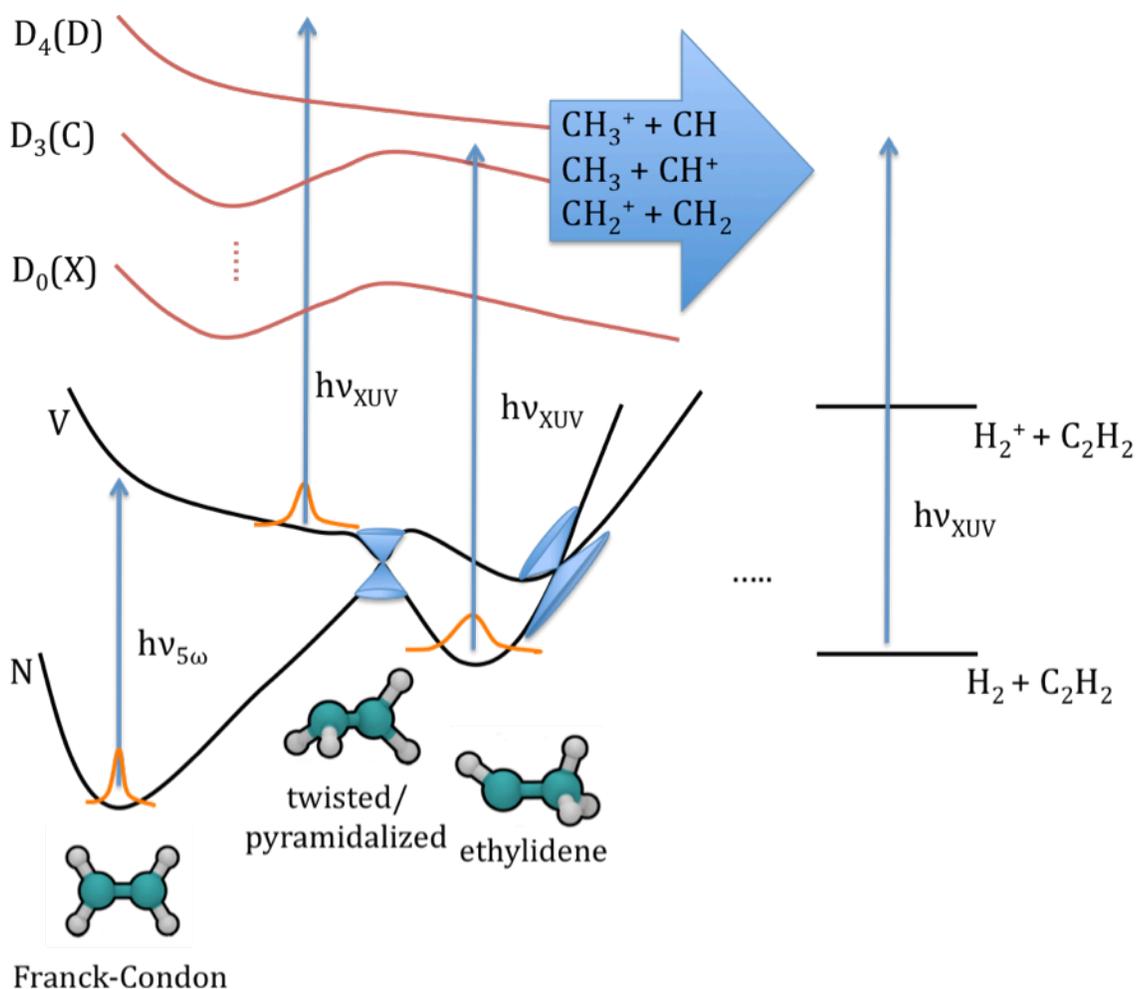

**Figure 1.** Schematic picture of the dynamics after excitation to the V state of ethylene. Two conical intersections are depicted, at twisted/pyramidalized and ethylidene geometries. Two instances of the XUV probe are also indicated, to a representative cation state. When the geometrical configuration of the molecule is $CH_2CH_2$, one can expect that many of these cation states will lead to symmetric fragmentation of the ion ($CH_2 + CH_2^+$). On the other hand, when the geometrical configuration of the molecule is more nearly ethylidene-like, one can expect that XUV excitation will lead to asymetric fragmentation of the ion ($CH + CH_3^+$). As depicted on the right, XUV absorption after $H_2$ elimination can also lead to ionization of $H_2$ (and analogously to ionization of H atoms, not shown). Thus, monitoring the dependence of ion fragments on pump-probe time delay provides a means to map out the dynamics of the excited state and potentially to determine whether and when each of the two types of intersections is involved in nonradiative decay.

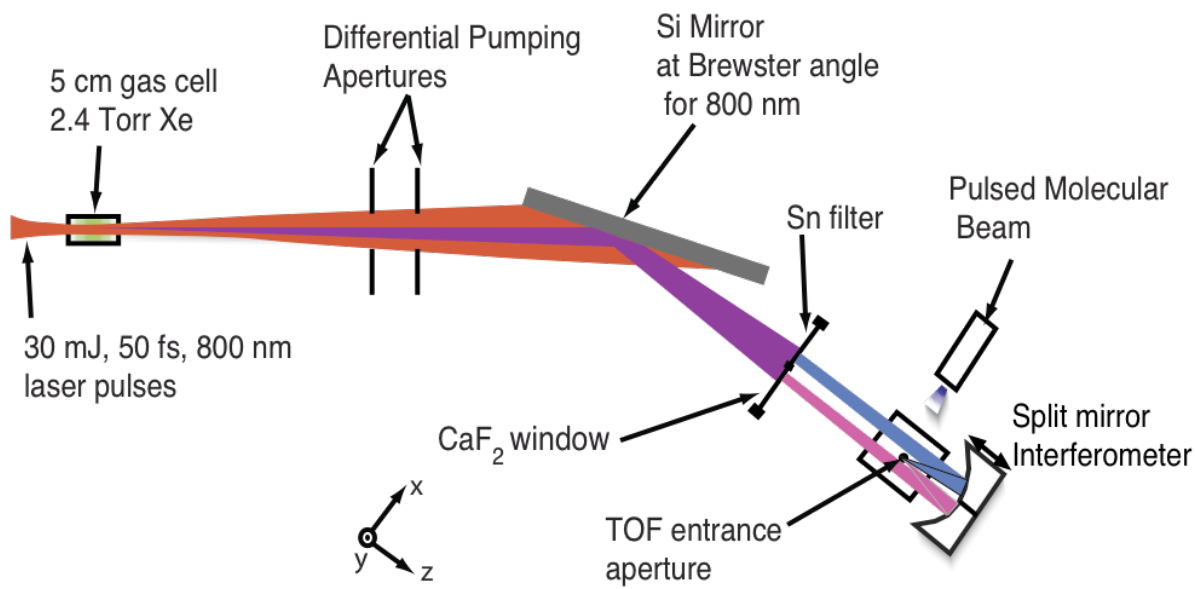

**Figure 2.** Pump/Probe Apparatus. High order harmonics are recombined with variable delay at the common focus of two spherical concave mirrors. Photo-ions are extracted from the focal region through a 500 micron pinhole with a strong electric field in the y direction.

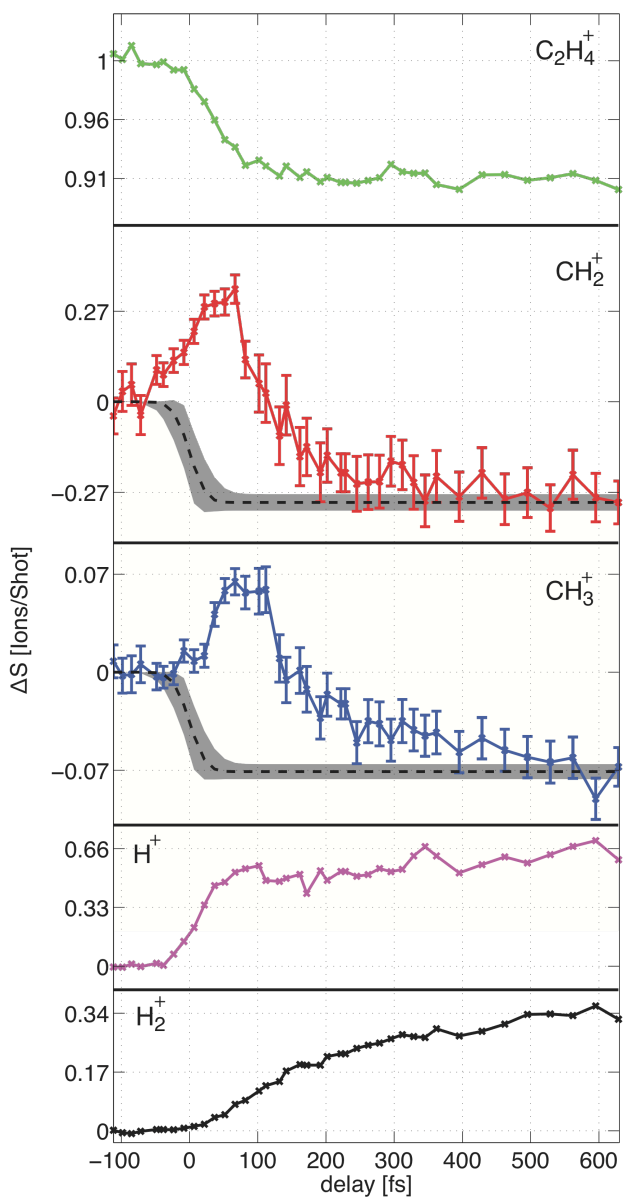

**Figure 3.** Absolute change in photo-ion yields in ions/shot for fragments $CH_2^+$, $CH_3^+$, $H^+$ and $H_2^+$, and corresponding relative change in percent for $C_2H_4^+$. Positive delay corresponds to the 7.7 eV probe pulse arriving first. Note the different y-scale for each fragment. The dashed black lines indicate the ground state bleach components present in the $CH_3^+$ and $CH_2^+$, with their associated systematic uncertainties present in the shaded grey regions.

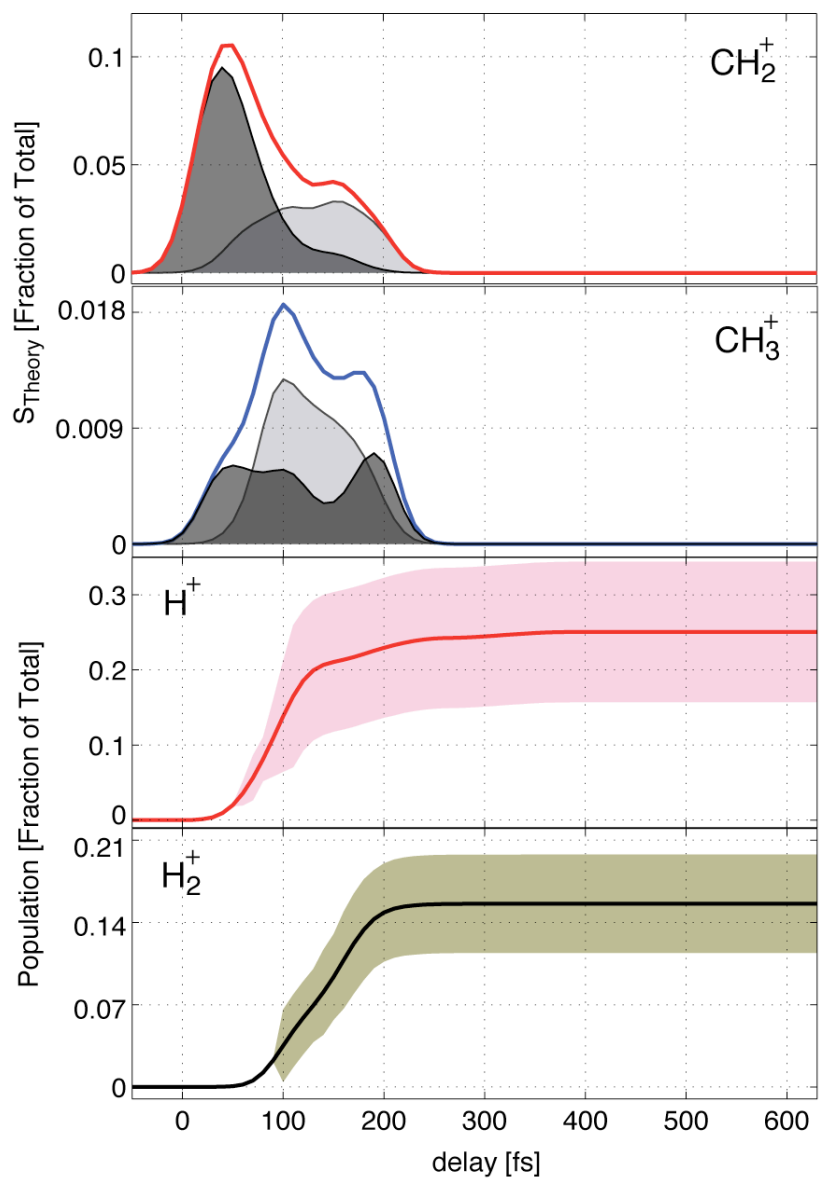

**Figure 4.** $CH_2^+$ and $CH_3^+$ panes (upper two panels) show simulated $CH_3^+$ and $CH_2^+$ ion fragment yields as a function of pump/probe time delay for excited ethylene molecules, reported as a fraction of the total population initially launched on the V state surface. The solid line corresponds to the total yield. The filled areas show the contributions from photoionization of population on the excited V state surface (dark gray) and the ground N state surface (light gray). The lower two panels show the cumulative fraction of the initial population that has undergone H or $H_2$ elimination. The shaded regions represent $1\sigma$ statistical uncertainty. Positive delay corresponds to the 7.7 eV pump pulse arriving first.

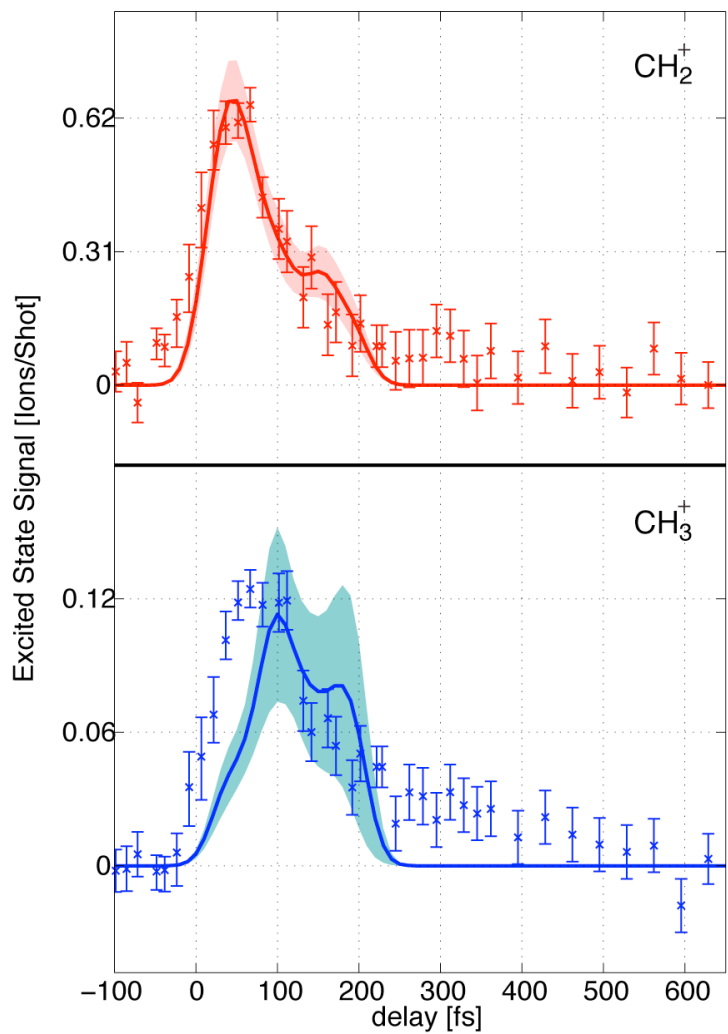

**Figure 5.** Comparison between theory and experiment for $CH_2^+$ (upper panel) and $CH_3^+$ (lower panel) fragments. Data Points: Excited molecule photoionization signals extracted from the raw data as described in the text. Solid Lines: $CH_2$ and $CH_3$ theory curves scaled by a common scale factor. Shaded regions represent the $1\sigma$ statistical uncertainty in the theoretical results.

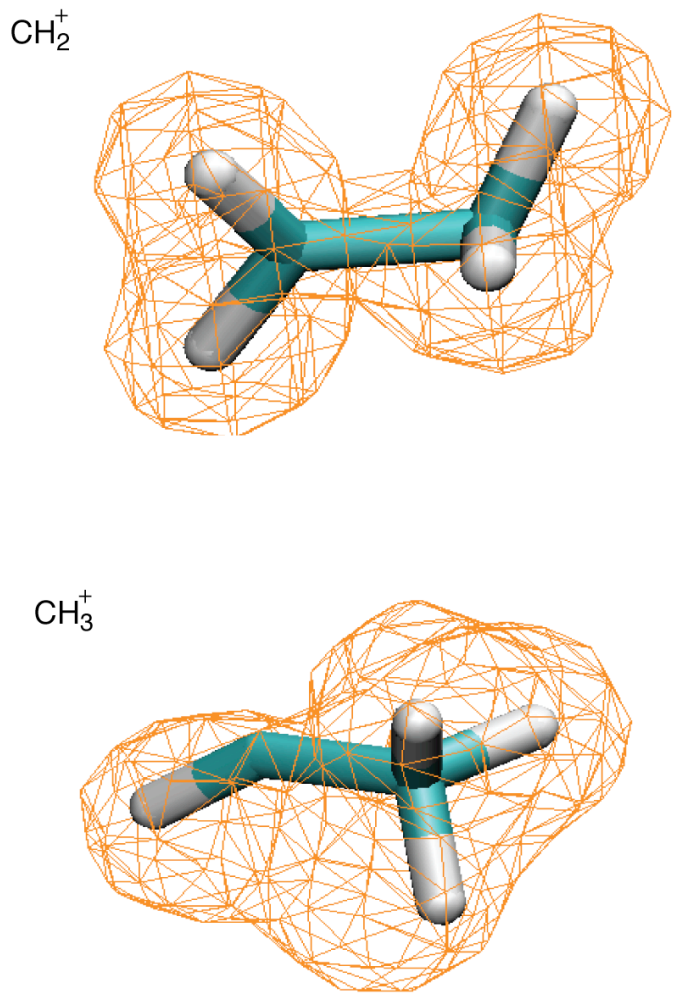

**Figure 6.** Density plot (created using volmap tool of VMD[47]) for trajectory basis functions that undergo direct (within 200fs) dissociation after XUV photo-ionization to give $CH_2^+$ (upper panel) or $CH_3^+$ (lower panel) fragments. The density plots are overlaid on the twisted/pyramidalized (upper panel) and ethylidene (lower panel) MECIs, showing that the experimental $CH_2^+/CH_3^+$ signals are strongly correlated with the two different MECIs that are operative in excited state quenching for ethylene.